\documentclass[twocolumn,aps,superscriptaddress]{revtex4-1}
\usepackage{palatino}
\usepackage[latin1]{inputenc}
\usepackage{epsf}
\usepackage{amsmath,amssymb}
\usepackage{latexsym}
\usepackage{calc}
\usepackage{xcolor}
\usepackage{graphicx}
\usepackage{hyperref}

\newcommand{\ben}{\begin{equation}}
\newcommand{\een}{\end{equation}}
\newcommand{\bea}{\begin{eqnarray}}
\newcommand{\eea}{\end{eqnarray}}

\def\sss{\scriptscriptstyle\rm}

\def\1s{_{1,\sss S}}
\def\2s{_{2,\sss S}}

\def\br{{\bf r}}
\def\bR{{\bf R}}

\def\bq{{\bf q}}

\def\dulr{{\underline{\underline{\bf r}}}}
\def\dulR{{\underline{\underline{\bf R}}}}

\begin{document}
\title{Exact-Factorization-Based Surface-Hopping for Multi-State Dynamics}
\author{Patricia Vindel-Zandbergen}
\affiliation{Department of Physics, Rutgers University, Newark, New Jersey 07102, USA}
\email{pv.zandbergen@rutgers.edu; neepa.maitra@rutgers.edu}
\author{Spiridoula Matsika}
\affiliation{Department of Chemistry, Temple University, Philadelphia, Pennsylvania 19122, United States}
\author{Neepa T. Maitra}
\affiliation{Department of Physics, Rutgers University, Newark, New Jersey  07102, USA}
\email{neepa.maitra@rutgers.edu}

%
%
%

\begin{abstract}
A surface-hopping algorithm recently derived from the exact factorization approach, SHXF, [Ha, Lee, Min, J. Phys. Chem. Lett. {\bf 9}, 1097 (2018)] introduces an additional term in the electronic equation of surface-hopping, which couples electronic states through the quantum momentum. This term not only provides a first-principles description of decoherence but here we show it is crucial to accurately capture non-adiabatic dynamics when more than two states are occupied at any given time. Using a vibronic coupling model of the uracil cation, we show that the lack of this term in traditional surface-hopping methods, including those with decoherence-corrections, leads to failure to predict the dynamics through a three-state intersection, while SHXF performs similarly to the multi-configuration time-dependent Hartree quantum dynamics benchmark. 
 \end{abstract}

\maketitle

The simulation of ultrafast dynamics after photo-excitation or when driven by a laser field involves electronic and nuclear motions that are highly correlated with each other.  Going beyond the Born-Oppenheimer (BO) approximation is essential yet challenging for more than a few degrees of freedom,  given that 
solving the molecular time-dependent Schr\"odinger equation scales exponentially with system-size.
Theoretical modeling of coupled electron-nuclear dynamics inevitably requires approximations, and most often a mixed quantum-classical method is used where the nuclei are treated via an ensemble of classical trajectories while a quantum mechanical description is retained for the electrons. 
 In this category the standard methods are Ehrenfest dynamics~\cite{T98,SNM85,MM80} and trajectory surface hopping (SH) dynamics~\cite{T90,LAP16,CB18,SJLP16}. 
 While Ehrenfest dynamics can be derived from a time-dependent self-consistent field ansatz, there is no first-principles derivation for SH, but it is generally preferred to Ehrenfest because of its ability to capture effects like wavepacket branching after passage through non-adiabatic coupling regions. 
 The two methods share the same electronic equation but differ in the classical force driving the nuclear motion. 
What the correct classical force {\it should} be in a mixed quantum-classical approach was recently resolved by the exact factorization approach~\cite{AMG10,AMG12,AASMMG15}. This approach enables the definition of exact time-dependent potentials driving the nuclear wavefunction that fully incorporate the coupling to the electrons, so the force obtained from considering them as classical potentials is the correct one in a mixed quantum-classical treatment. 

A self-consistent mixed quantum-classical method derived from the exact factorization is the coupled-trajectory mixed quantum-classical scheme (CTMQC)~\cite{MAG15,AMAG16,MATG17,GAM18,AG21,PA21,AC19}. The equations for the electronic system and the nuclear trajectories differ from both Ehrenfest and SH, with 
additional terms that couple the classical trajectories via the nuclear quantum momentum.
 Importantly, the additional term in the electronic equation was shown to induce decoherence after passage through a non-adiabatic coupling region~\cite{AMAG16,GAM18}, nudging the electronic coefficients associated with a given trajectory to be non-zero only on one state. This property is especially enticing for SH schemes, where the disconnect between the coherent electronic evolution propagating in a superposition of states and the nuclear trajectory evolving on a single state at any given time leads to the problem commonly known as ``overcoherence". Thus the exact factorization-based surface-hopping scheme (SHXF) was introduced, where the electronic equation from CTMQC is used in the SH algorithm~\cite{HLM18,PyUNIxMD} and restores the internal consistency, incorporating decoherence from first-principles. This method has been demonstrated on a range of fascinating light-induced processes on complex molecules~\cite{FPMC19,FPMK18,FMK19,FMC19}. Its performance was recently benchmarked against the high-level ab initio multiple spawning method~\cite{BQM00,BM98,CM18} on some small molecules, and compared with the more {\it ad hoc} decoherence corrections introduced previously~\cite{VIHMCM21}. The form of the first-principles decoherence correction and its effect on individual trajectories was found to be qualitatively different than the previous corrections, whose effects also differed from each other, yet averages over trajectories were similar for these molecules.

  In this Letter, we show that the effect of the electronic equation from the exact factorization approach goes far beyond providing a first-principles decoherence correction: it  induces electronic transitions mediated by the nuclear quantum momentum which, in the SHXF scheme, significantly affect hopping probabilities and resulting dynamics when more than two states become occupied at any given time. 
 This is important for three-state and four-state conical intersections, which have been realized to be abundant in polytatomic molecules~\cite{M05,MY02,MY03,M08,CM05,KM08}, affecting photochemical pathways~\cite{AWG12,DCG10}. 
  These transitions are completely missed by conventional SH (and Ehrenfest) methods, including those with decoherence corrections, resulting in erroneous dynamics. We demonstrate this on dynamics through a three-state intersection in the photo-excited uracil cation, which SHXF captures accurately, while the other decoherence-corrected SH schemes fail.

In SH methods, an ensemble of independent classical nuclear trajectories evolve on a single BO potential energy surface at any given time, making hops between them according to a stochastic algorithm~\cite{T90,LAP16,CB18,SJLP16}. Labelling each nucleus by $\nu$, each classical trajectory, ${\bf R}_\nu^{(J)}(t)$, is associated with an electronic state, which is expanded in the BO basis,  $\Phi^{(J)}(\dulr, t) = \sum_n C_n^{(J)}(t) \Phi^{{\rm BO}}_{{\dulR^{(J)}(t)},n}(\dulr, t)$, with coefficients satisfying 
\ben
\dot C_n^{(J)} = -\frac{i}{\hbar}\epsilon_n^{(J)}C_n^{(J)} -\sum_k \sum_\nu {\bf d}^{(J)}_{nk,\nu}\cdot{\dot{\bf R}^{(J)}_\nu}C_k^{(J)} + \xi^{(J)}_n
\label{eq:SH}
\een
where $\epsilon_n^{(J)}= \epsilon_n(\dulR^{(J)}(t))$ is the BO potential energy surface evaluated at the current position of the nuclear trajectory and ${\bf d}^{(J)}_{nk,\nu} = \left.\langle \Phi^{\rm BO}_{n}\vert\nabla_\nu\Phi^{\rm BO}_{k} \rangle\right\vert_{\dulR^{(J)}(t)}$ is the non-adiabatic coupling vector between BO states $n$ and $k$. (The notation $\dulR$ indicates all nuclear coordinates). 
The time-dependence of each term in Eq.~(\ref{eq:SH}) is omitted for notational simplicity.  The original SH algorithm involves just the first two terms on the right in Eq.~(\ref{eq:SH}), while in 
SHXF~\cite{HLM18,PyUNIxMD}, which uses the electronic equation coming from the mixed quantum-classical limit of the exact factorization approach, we have
\ben
\xi^{(J)}_n =  \sum_k \sum_\nu\frac{1}{M_\nu}\left.\frac{\nabla_\nu |\chi|}{\vert\chi\vert}\right\vert_{\dulR^{(J)}(t)}\cdot
\left({\bf f}^{(J)}_{k,\nu} - {\bf f}^{(J)}_{n,\nu}\right) \vert C_k^{(J)} \vert^2 C_n^{(J)}
\label{eq:SHXFe}
\een
where $\vert \chi(\bR,t)\vert^2$ is the nuclear density reconstructed from the classical trajectories, and ${\bf f}^{(J)}_{k,\nu} = -\int^t\nabla_\nu \epsilon_{k}^{(J)}(t') dt'$ is an effective state-dependent momentum computed from the  force from the $k$th BO surface integrated along the trajectory. This term introduces couplings between electronic states that are occupied at time $t$, mediated by the nuclear quantum momentum $\frac{\nabla_\nu |\chi|}{\vert\chi\vert}$ and the difference between the effective state-dependent momenta. The nuclear quantum momentum introduces an effective coupling between neighboring nuclear trajectories, determined by the slope of the nuclear distribution at the given trajectory. 
  Refs.~\cite{AMAG16,GAM18} give details on the mechanics of how this term leads to decoherence and wavepacket splitting in model systems. In the SHXF algorithm, these forces and momenta are cleverly calculated via the use of auxiliary trajectories  launched on surfaces when the electronic coefficient of that state becomes higher than a small threshold; this enables the procedure to stay within the computationally efficient independent trajectory picture. 
 The reader is referred to Refs.~\cite{HLM18} for details on the formalism and algorithm. Along with other MQC approaches for correlated electron-nuclear dynamics simulations, SHXF is implemented in the molecular dynamics package, PyUNIxMD~\cite{PyUNIxMD}. 



\begin{figure}[!htbp]
	\includegraphics[width=0.5\textwidth]{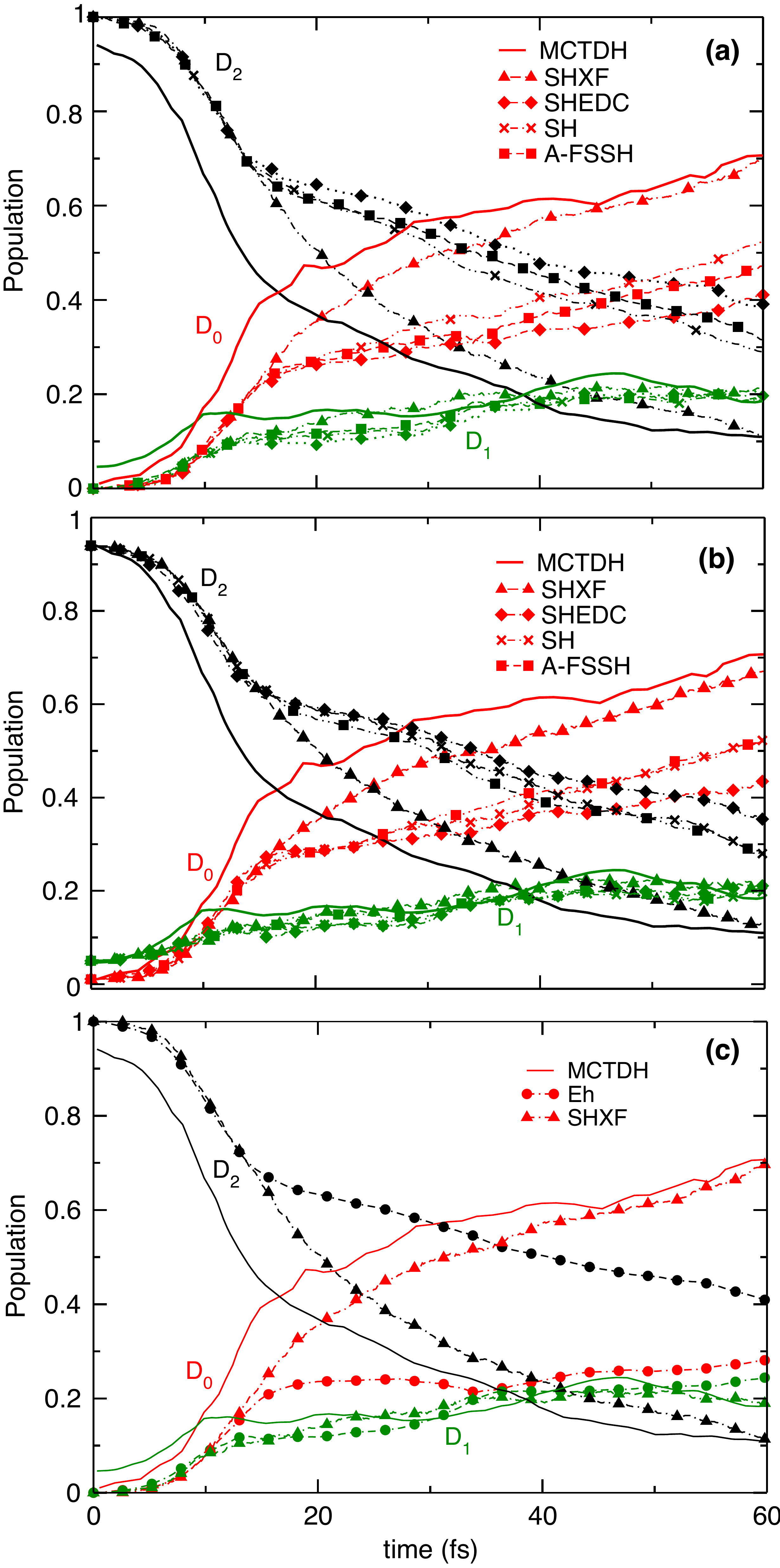}
\caption{Population dynamics in the uracil cation: a) SHXF, SH, SHEDC and A-FSSH, all with momentum adjustment along the NACV, and beginning in the adiabatic D$_2$ state, along with the reference MCTDH  populations (from Ref.~\cite{AKM15}) that begins in the diabatic state. Black, green and red lines correspond to D$_2$,D$_1$ and D$_0$ states respectively. b) As above but with the SHXF, SH, SHEDC, and A-FSSH calculations all beginning in the mixed state with 94\% population in D$_2$, 5\% in D$_1$ and 1\% in D$_0$. c) Population dynamics calculated from Ehrenfest and SHXF simulations starting in the adiabatic D$_2$ state, along with the reference MCTDH. }
\label{fig_Uracil_1}
\end{figure}

Recently, the SHXF approach was compared with 
two widely-used decoherence corrected SH methods~\cite{VIHMCM21}, the energy-based decoherence correction (SHEDC)~\cite{GP07,GPZ2010} and the augmented fewest switches surface-hopping (A-FSSH)~\cite{SOL13,JES16}. Their decoherence-corrections can be represented by a term $\xi^{(J)}_n$ in Eq.~(\ref{eq:SH}), for which the three methods are very different from each other,  resulting in remarkably different decoherence mechanisms at the individual trajectory level~\cite{VIHMCM21}. The
SHEDC correction  is imposed as an exponential decay on non-active states to damp their amplitudes at a rate that depends on their potential energy difference with the active state, 
and the kinetic energy\cite{GP07,GPZ2010,ZNJT04,ZJT05}. 
Instead, A-FSSH exploits the use of auxiliary trajectories evolving classically on different surfaces, defining a decoherence time through how fast they deviate from the active surface, used to collapse non-active state coefficients~\cite{JES16} through a stochastic procedure.

There are significant differences between the $\xi_n$  derived from first-principles and that from the somewhat {\it ad hoc} SHEDC or A-FSSH methods. First, is that the latter are introduced as decoherence rates, acting on only the coefficients of the non-active states; in contrast, the non-linear dependence on the coefficients of the first-principles $\xi_n$ gives a structure more complex than  a simple exponential decay rate. Further, it acts on all coefficients including the active state. Most crucial for the present work, is that with the first-principles-derived $\xi_n$, 
 each state is coupled to {\it all} occupied states, while in SHEDC or A-FSSH it couples a non-active state $n$ only to the active state, which means that when more than two states are occupied at a given time, this will yield fundamentally different electronic coefficients over time. This leads to significantly different hopping probabilities and resulting dynamics than uncorrected SH, SHEDC, or A-FSSH, which we expect to be particularly notable in problems such as three-state conical intersections, accessible after UV-excitation in some DNA and RNA bases, as we will demonstrate shortly, and in general radicals involving Rydberg states~\cite{M05,MY02,MY03}.
Out of other decoherence-corrected SH schemes, one that shares the non-linearity feature and the dependence on the coefficients of all states, is the decoherence-induced SH (DISH) algorithm~\cite{AP14,JFP12}; but, similar to other {\it ad hoc} decoherence-corrected SH schemes, the correction is viewed as a collapse-rate probability rather than a coherent contribution to the time-evolution of the coefficients.

\begin{figure}[h]
\includegraphics[width=0.5\textwidth,height=0.45\textwidth]{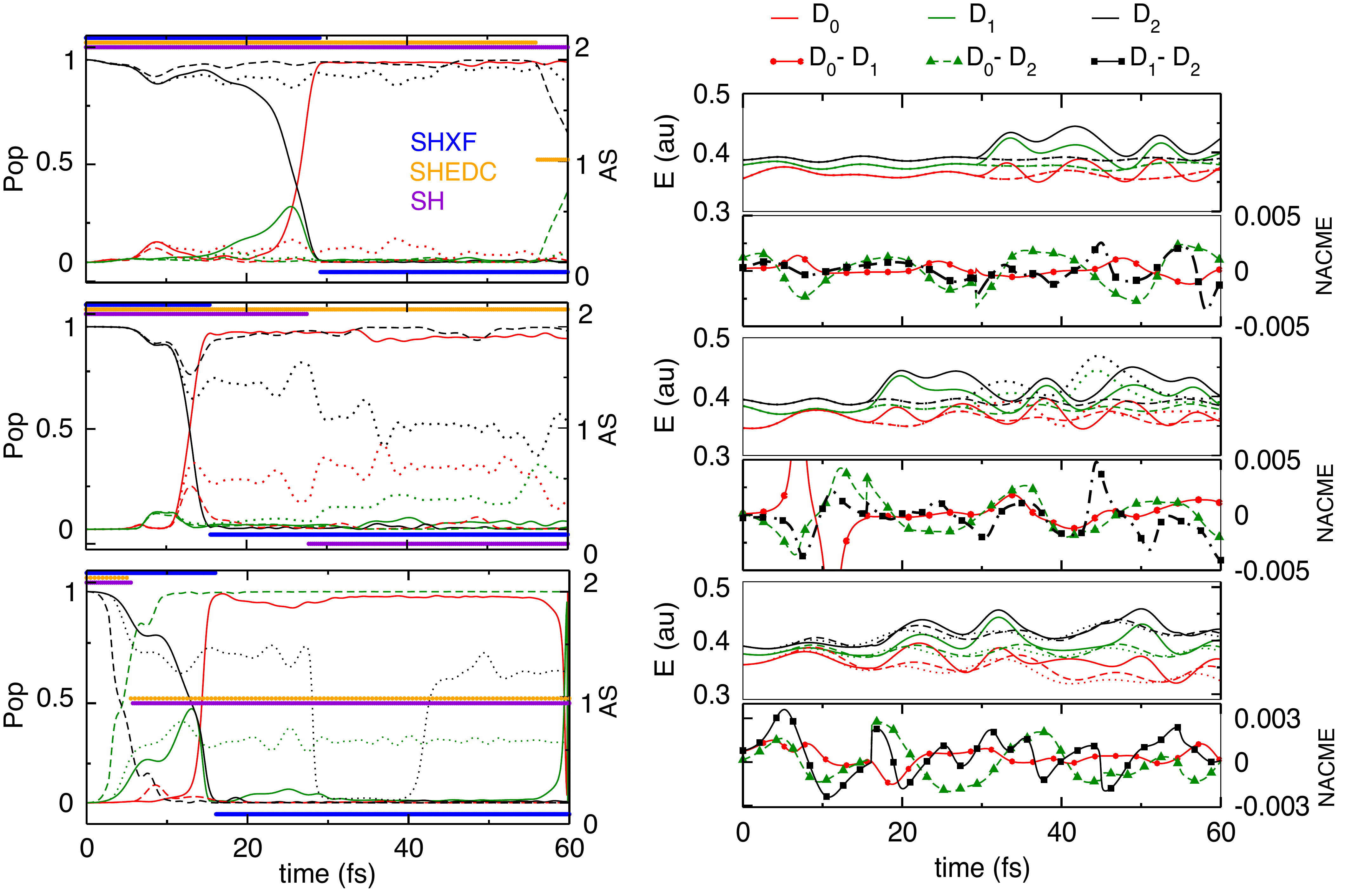}
	\caption{Left panels: Population dynamics in the uracil cation for three trajectories with the same initial conditions: SH, SHEDC and SHXF. Black, green and red lines correspond to the D$_2$, D$_1$ and D$_0$ electronic populations, respectively, with solid lines for SHXF, dashed for SHEDC and dotted lines for SH. The colored symbols indicate the active state (AS): blue for SHXF, orange for SHEDC and purple for SH. Right panels show the electronic energies (upper) and NACMEs between pairs of states (lower) during SHXF dynamics.}
\label{fig_Uracil_2}
\end{figure}

We now demonstrate the impact of this additional term in SHXF 
on the relaxation dynamics of the uracil radical cation after photo-ionization, and compare our results with SH, SHEDC, A-FSSH, and Ehrenfest simulations, using quantum dynamics from  multi-configuration time-dependent Hartree (MCTDH) obtained in Refs.~\cite{AKM15,AWM16} as the reference. 
The excited states of the uracil cation have been thoroughly studied~\cite{M08,AKM15,AWM16,STB19}: a three-state and a two-state conical intersection (CI) between D$_0$/D$_1$/D$_2$ and D$_0$/D$_1$ states, respectively, were found, with minima located 0.95 eV and 0.45 eV respectively, above the D$_0$ minimum.
We model the initial photo-ionized state by placing the ground nuclear wavepacket at the $S_0$ minimum of the neutral on the vertically-excited  D$_2$ excited cationic state. 
This state relaxes to the ground-state before fragmentation occurs, and MCTDH calculations of Ref.~\cite{AKM15} indicate that the 3-state and 2-state CIs are crucial in the transition to the ground-state on the femtosecond time-scale.

These MCTDH calculations were performed on an eight-mode linear vibronic coupling (LVC) model developed in Ref.~\cite{AKM15}, coupling the 4 lowest states of the cation; the model was validated by comparing the MCTDH autocorrelation spectrum with experiment. To make a meaningful comparison, we use the same vibronic coupling Hamiltonian  for the SH dynamics. The normal modes included in the model along with the parameters of the model are provided in the Supporting Information. 
Since the LVC model breaks down for too large excursions of the normal mode coordinates, we limit the duration of our simulations to 60 fs.


The SH, SHEDC, SHXF, and Ehrenfest calculations are performed using the PyUNIxMD code~\cite{PyUNIxMD}, while A-FSSH computations are done with the  SHARC 2.0 code~\cite{SHARC,MMG18,sharc_prog}. The momentum is rescaled  after a hop in the direction of the non-adiabatic coupling vector (NACV), and after a frustrated hop the momentum direction is preserved, unless otherwise stated. Convergence was achieved with a nuclear time-step of $dt = 0.1$fs for SHXF and $dt=0.25$fs for SHEDC and A-FSSH and $N = 1000$ trajectories. The initial nuclear coordinates and momenta are obtained from Wigner-sampling the ground-state at the $S_0$ minimum using the frequencies of the eight modes in the model, and the initial electronic state is D$_2$, unless otherwise stated.
In computing the quantum momentum from the auxiliary trajectories a width of $\sigma = 0.08$a.u. is used, determined from the initial distribution of the nuclear trajectories of the C=C, C=O and C-N bonds.
Further details on the calculations are included in the Supporting Information.

Figure~\ref{fig_Uracil_1} shows the population dynamics as the fraction of trajectories running on each state, $\Pi_k(t) = N_k(t)/N$, computed from the SH, SHEDC, A-FSSH and SHXF simulations with MCTDH as a reference. In the top panel, all the surface-hopping calculations begin in the adiabatic D$_2$ state.  
There is a small initial discrepancy with the reference because the MCTDH calculations of Refs.~\cite{AKM15,AWM16} began in the diabatic state, which has a 0.94 projection on to the adiabatic D$_2$ state. So in the middle panel,  we instead begin the surface-hopping calculations in a mixed state: from the 1000 initial conditions, $940$ trajectories start in the D$_2$ state, $50$ in D$_1$ and $10$ in D$_0$, such that the corresponding adiabatic populations are close to those of the diabatic state that the MCTDH began in. ~\footnote{We note that  this is not the same as starting in the diabatic state, since it represents an incoherent mixture rather than a coherent initial state, however it approximates the initial populations of the reference calculation more accurately than starting in the adiabatic state.} Starting in this initial mixed state does not significantly change the population traces from those starting in the adiabatic state, and certainly does not change the most salient observation:
 the SHXF population traces follow that of the MCTDH much more closely than the other SH methods do. 
We observe that all the methods predict the same initial decay of the $D_2$ state and agree with the MCTDH reference until about 15 fs. Importantly, the population in D$_1$ and D$_0$ states begin to rise simultaneously, suggesting that the molecule has reached the 3-state CI seam with population transfer occurring from D$_2\rightarrow$ D$_1$ and also directly from D$_2\rightarrow$ D$_0$. After about 15 fs, there is a remarkable difference in the population traces between SHXF and all the other SH calculations. The rate of decay of the D$_2$ population in SHXF along with the increase of D$_0$ closely follow that predicted by MCTDH, while SH, SHEDC, and A-FSSH show a qualitatively different trend, but similar to each other, with a distinctly slower rate of decay at around 15 fs in D$_2$ and a much gentler rise in D$_0$ population. Interestingly, all methods predict similar population  dynamics for the D$_1$ state. SH, SHEDC, and A-FSSH methods all significantly underestimate the D$_2\rightarrow$  D$_0$ transitions, in contrast to SHXF which shows a far closer agreement with the reference MCTDH calculation. 
The lowest panel of Figure~\ref{fig_Uracil_1} shows the Ehrenfest dynamics, which performs worse than the standard SH methods. 

While the traditional decoherence-corrected methods SHEDC and A-FSSH hardly change the populations determined by the fraction of trajectories $\Pi_k(t)$ predicted by the original uncorrected SH method, SHEDC does correct the internal consistency of SH, i.e. the agreement of electronic populations $\vert C_k(t)\vert^2$ with  $\Pi_k(t)$. Figure S1 in the Supporting Information compares $\vert C_k(t)\vert^2$ and  $\Pi_k(t)$ for the different methods. Surprisingly, A-FSSH  does not have any noticeable effect and its internal consistency error is comparable to uncorrected SH.  Internal consistency is satisfied well in SHXF as shown in that figure. 




To understand why SHXF captures the dynamics accurately while SH, SHEDC, and A-FSSH do not, we first note that SHXF has almost twice as many D$_2\rightarrow$  D$_0$ hops than the three other SH methods. A figure showing the number of hops between the different pairs of states is provided in the Supporting Information. Yet, the hopping probability is determined by the same fewest-switches stochastic algorithm: a hop is made from active state $a$ to state $n$ if $\Sigma_{k=1}^{n-1} \zeta_{ak} < r \le \Sigma_{k=1}^{n} \zeta_{ak}$ where $r$ is a random number uniformly distributed in $[0,1]$ and
 $\zeta_{ak} = {\rm max}\left( 0, -2 Re (C_a^*C_k d_{ka,\nu}\cdot \dot{\bR}_\nu dt/\vert C_a\vert^2\right)$. This means that the coefficients $C_0$ and $C_2$ must have evolved to be significantly different in SHXF than those from SH, SHEDC, and A-FSSH, due to the third term $\xi_n$ in Eq.~(\ref{eq:SH}). As noted earlier, this term introduces a distinctly different dynamics to the coefficients than in the other methods, due in particular to its coupling of all occupied states and its nonlinearity.  The term is also lacking in the Ehrenfest calculation which has the same electronic equation as in SH.

To see its effect in practise, we take a look at the individual trajectory level. In Figure \ref{fig_Uracil_2} we show the populations and the active states for three randomly chosen trajectories in the SH, SHEDC, and SHXF simulations, along with the BO energies of each state and the non-adiabatic matrix elements (NACMEs) for the SHXF trajectories.
What had been noted for the molecules studied in Ref.~\cite{VIHMCM21}, was that the different structure of the corrections lead to a  distinct behavior at the individual trajectory level, with the SHEDC correction dampening the populations of non-active states in a largely monotonic way, while SHXF showing a more oscillatory behavior before a complete decoherence.  The dynamics in those molecules were essentially two-state dynamics, while for our LVC model of the uracil cation here, we encounter regions where three states are occupied at a given time. In SHXF, all three states enter into the evolution of any coefficient through the $\xi_n$ of Eq.~(\ref{eq:SH})--(\ref{eq:SHXFe}), while in SHEDC and A-FSSH this term involves only the state in question and the active state 
(and this term is zero in SH).  In Fig.~\ref{fig_Uracil_2} we clearly see the effect of this additional coupling: when the coefficients of the D$_0$, D$_1$ and D$_2$ states begin to become non-zero from the non-adiabatic coupling term kicking in,  the SHXF has a complex dynamics that enhances  population transfer to inactive states, leading ultimately to a larger $\vert C_0\vert$ that  increases the D$_2 \rightarrow$ D$_0$ hopping probability. 
The NACMEs themselves for  D$_2 \rightarrow$ D$_0$  and  D$_2 \rightarrow$ D$_1$ are of similar magnitude along the trajectories (the smaller D$_2$--D$_1$ energy spacing is (more than) compensated by the larger D$_2$--D$_0$ energy-gradient coupling matrix element~\cite{AWM16}). 
Away from the intersection region, the term decoheres the SHXF populations, as observed in the earlier studies. 
On the other hand, throughout the dynamics, SHEDC quickly dampens the populations of non-active states to zero, and a hop generally does not occur unless the couplings are very large.  The original uncorrected SH propagates using only the first two terms of Eq.~(\ref{eq:SH}), lacking both decoherence as well as the additional coherence from coupling.  


We finally note that a SHEDC calculation on the full-dimensional uracil cation was performed in Ref.~\cite{AWM16}, and although the overall population trend was somewhat similar to that of the MCTDH in the eight-mode LVC model, the initial behavior was quite different; in particular,  the full-dimensional SHEDC calculation does not show an initial simultaneous rise of D$_1$ and D$_0$ as observed in the eight-mode LVC model here, suggesting that the three-state intersection region is not probed as closely. 
 In the Supporting Information, we provide a comparison between the full-dimensional and LVC SHEDC and SHXF calculations for the populations and bond-lengths. 
 
 In summary, the electronic equation derived from a mixed quantum-classical limit of the exact factorization induces electronic transitions in a surface-hopping scheme that leads to strikingly different dynamics than conventional (decoherence-corrected) SH schemes when several electronic states become occupied at a given time. This gives dramatically improved agreement with reference quantum dynamics calculations.  The exact-factorization-based approach yields an additional term to the original SH algorithm, that couples all occupied states through the nuclear quantum momentum, and leads to coherent couplings near interaction regions as well as  decoherence away from them. The implementation of the SHXF algorithm of Ref.~\cite{HLM18,PyUNIxMD}  through auxiliary trajectories launched for each independent trajectory is a computationally efficient approach that will likely lead to improved predictions of surface-hopping dynamics
 for complex photo-excited dynamics in molecules beyond two-state dynamics, and could also be used in conjunction with a quantum trajectory SH scheme where {\it ad hoc} velocity adjustment is not required~\cite{M19b}. 
 It is also likely to be promising for situations where multiple non-adiabatic events occur, including laser-driven processes and polaritonic chemistry. The SHXF approach does not capture explicit nuclear phase effects, but whether these are important or not in the XF representation of dynamics also merits further study~\cite{CAG16,CA17,AC18}.

\begin{acknowledgements}
This work was supported by the Computational Chemistry Center: Chemistry in Solution and at Interfaces funded by the U.S. Department of Energy, Office of Science Basic Energy Sciences, under award no. DE-SC0019394  (P.V.Z. and S.M.) as part of the Computational Chemical Sciences Program. 
Partial support from the U.S. Department of Energy, Office
of Basic Energy Sciences, Division of Chemical Sciences,
Geosciences and Biosciences under Award DE-SC0020044 (N.T.M.),  
are gratefully acknowledged. P.V.Z. acknowledges partial support provided by Molecular Sciences Software Institute (MolSSI) through the Seed Software Fellowship under the NSF grant 1547580. This research used resources of the National Energy Research Scientific Computing Center, a DOE Office of Science User Facility supported by the Office of Science of the U.S. Department of Energy under Contract No. DE-AC02-05CH11231.
\end{acknowledgements}


\newpage
\section{Supporting Information: Vibronic coupling model}
The vibronic coupling model we use is taken from Ref.~\cite{AKM15}, based on the framework of Refs~\cite{KDC84,DYK04,DYK11}. 
 Out of the models presented there, we select the eight-mode model of 6 modes of $a'$ symmetry that connect states of the same symmetry, and 2 modes of $a''$ symmetry that connect neighboring states of different symmetry. The 6 $a'$ modes correspond to C$=$O stretch (labelled as modes 25 and 26 in Ref.~\cite{AKM15}), CN stretch plus NCH/CNH bending (modes 18, 20, 21) and C$=$C stretch (mode 24). The $a''$ modes  (labelled as 10 and 12) are out-of-plane motions. 

The diabatic potential energy surfaces (PES) are expressed in terms of the mass-weighted vibrational mode coordinates $\bq$ around a reference geometry $\bq_0$ and the Hamiltonian matrix is then written as:
\begin{equation}
	{\bf {H}} = {\bf {H}}^{(0)}+ {\bf {W}}
\end{equation}
where ${\bf{H}}^{(0)}$ is the ``zeroth-order" Hamiltonian which is diagonal in the diabatic basis and has identical  $\bf{q}$-dependence for each electronic state, and ${\bf{W}}$ is the matrix that contains all the state-specific  and coupling terms. Thus the elements of ${\bf{H}}^{(0)}$ are
(in atomic units):
\begin{equation}
	H_{\alpha\alpha}^{(0)} = E_{\alpha}+\sum_i (T_i+V_i) = E_{\alpha}+\sum_i{\left(\frac{\omega_{i}}{2}\frac{\partial^2}{\partial q_i^2}+V(q_i)\right)}
	\label{eq_harm}
\end{equation}
where $\omega_i$ is the vibrational frequency for each mode $i$ and $E_{\alpha}$ is the energy of the electronic state $\alpha$ at $\bq ={\bq}_0$. The sum goes over the eight normal modes in the model.  As the model is developed for the uracil cation with $\bq_0$ as the geometry of the neutral S$_0$ minimum, the energies $E_\alpha$ actually correspond to the ionization potentials. 
The potential $V(q_i)$ is taken to be harmonic for modes 10, 12, 18, 20, and 21, while a Morse potential is used to account for anharmonicity in modes 24, 25, and 26:
\bea
V(q_i) &=& \omega_i q_i^2/2,\;\;\;\; \;\;\;\;\;\;\;\;\;\;\;\;\;{\rm modes}\; 10, 12, 18, 20, 21\\
\nonumber
V(q_i) &=&d_{0,i}^{(\alpha)}[e^{(a_i^{(\alpha)}(q_i-q_{i,0}^{(\alpha)}))}-1]^2+e_{0,i}^{(\alpha)}, \;\;\;  {\rm modes}\; 24, 25, 26\\
\eea
In the LVC, the ${\bf W}$ matrix contains diagonal linear and quadratic corrections as well as the first-order coupling terms between states:
\begin{equation}
	W_{\alpha\alpha}^{(1)}=\sum_i \kappa_i^{\alpha}q_i, \;\;\; W_{\alpha\alpha}^{(2)}= \frac{1}{2}\sum_i \gamma_i^{\alpha}q_i^2
\end{equation}

\begin{equation}
         W_{\alpha\beta}^{(1)}=\sum_i \lambda_i^{\alpha\beta}q_i, \;\alpha\ne\beta
\end{equation}
while for the $a''$ modes (mode 10 and 12) a quartic form was required to give a double-well structure:
\ben
W_{\alpha\alpha}^{(4)}=\frac{1}{24} \sum_i k_i^{(\alpha)}q_i^4
\een

Ref.~\cite{AKM15} obtained the PES and coupling parameters  by fitting the model to \textit{ab initio} electronic structure calculations.
The  parameters for the model are given in Tables~\ref{table_1}, \ref{table_2} and \ref{table_3}. It should be noted however, that in the dynamics simulations the parameters for the $a''$ modes enter only through the couplings; that is, in the dynamics the diagonal potential terms from the $a''$ contributions are put to zero. This is to be consistent with what was run in the MCTDH calculations of Ref.~\cite{AKM15,AWM16}.

Diagonalization of the resulting ${\bf H}$ gives the adiabatic PES in terms of 
the normal mode coordinates, $\bq$, while the computation of SH dynamics uses cartesian coordinates for geometries and velocities. The initial geometries and velocities are sampled using a Wigner distribution created from the frequencies of the eight normal modes included in the model. 
A conversion between the two sets of coordinates is needed at each time step. 
Each normal coordinate is defined as a dimensionless mass-frequency scaled normal coordinate and calculated from the cartesian coordinates as \cite{WC04,PMFGDG19,ZHPMG21}
\begin{equation}
q_{i}=\sqrt{\omega_i}\sum_n \mathrm{\bf{K}}_{ni}\sqrt{M_n}r_n
\end{equation}
where $M_{n}$ is the atomic mass, $r_n$ is the displacement in cartesian coordinates and {\bf K}$_{ni}$ denotes the orthogonal conversion matrix between mass-weighted cartesian and normal coordinates. The displacement vector $\br$ is calculated at each time step as the difference between the geometry at time $t$ and the $\bq_0$ geometry, that is, the S$_0$ geometry of the uracil cation in this work. The reader is referred to Ref.~\cite{PMFGDG19} and the SHARC 2.0 package for details on the equations and implementation.

\section{Supporting Information: Internal Consistency}
The traditional decoherence-corrected methods SHEDC and A-FSSH hardly change the populations determined by the fraction of trajectories $\Pi_k(t)$ predicted by the original uncorrected SH method,  as observed in the main text, raising the question of whether they have any effect at all. 
In Figure~\ref{fig_Uracil_2} we plot $\Pi_k(t)$ and the electronic populations $\vert C_k(t)\vert^2$ for  the D$_2$, D$_1$ and D$_0$ states calculated from the SH, SHEDC, A-FSSH and SHXF simulations; the deviation of $\Pi_k(t)$ from $\vert C_k(t)\vert^2$ indicates violation of internal consistency. We see that SHEDC corrects the internal consistency error of SH, but, surprisingly, A-FSSH  does not have any noticeable effect and its internal consistency error is comparable to uncorrected SH.  Internal consistency is well satisfied in SHXF, which corrects both $\Pi_k(t)$ and $\vert C_k(t)\vert^2$ populations towards the reference MCTDH.

\section{Supporting Information: Number of hops in SH dynamics}
\begin{figure}[h]
\includegraphics[width=0.5\textwidth]{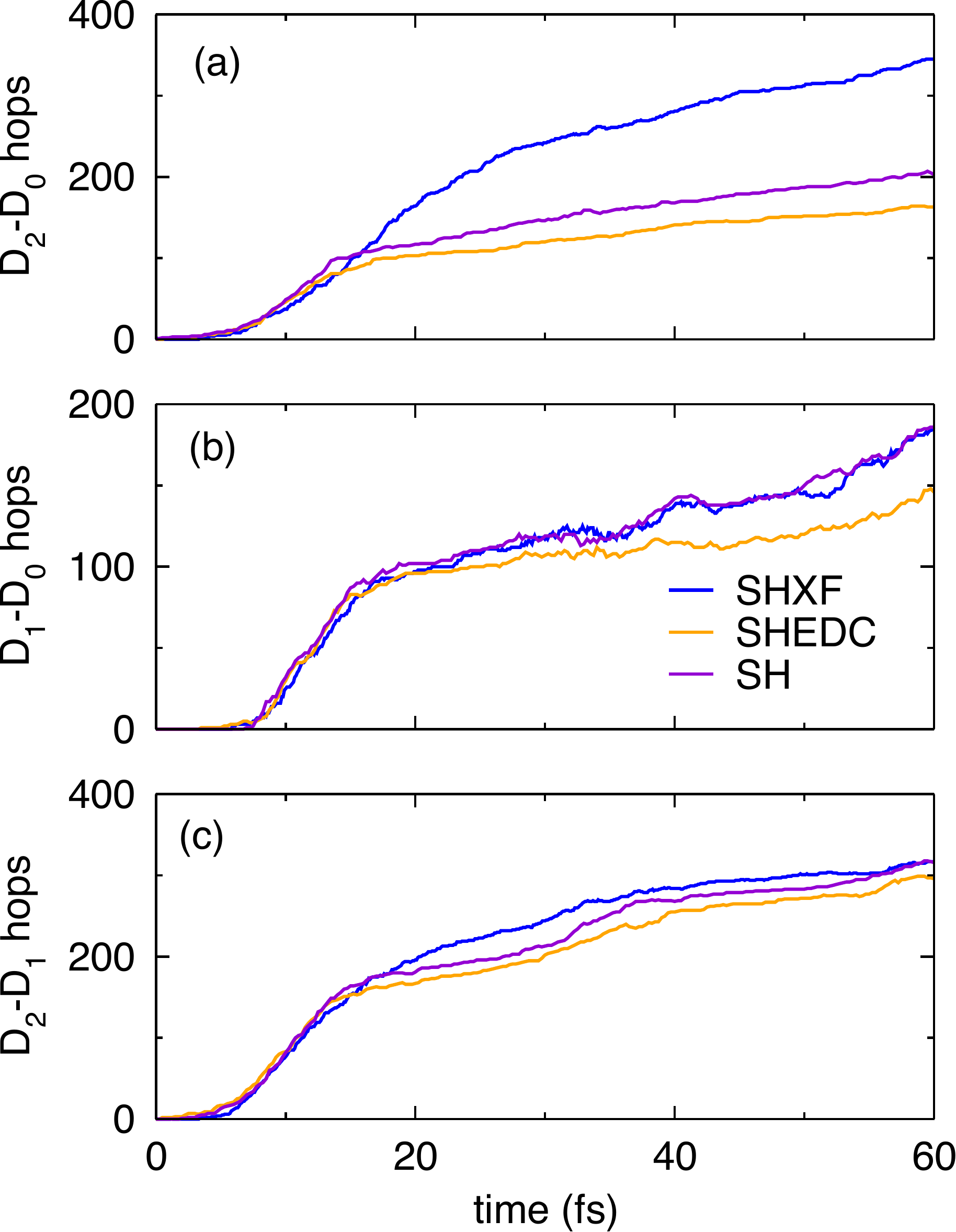}
	\caption{Number of hops between (a) D$_2$ and D$_0$, (b) D$_1$ and D$_0$ and (c) D$_2$ and D$_1$ in the uracil cation for SHXF (blue), SHEDC (orange) and SH (purple).}
\label{fig_hops}
\end{figure}

In Fig.\ref{fig_hops} we plot the accumulated number of hops between pairs of states (D$_2$-D$_0$, D$_1$-D$_0$, and D$_2$-D$_1$) from SH, SHXF and SHEDC simulations. The net hops are calculated as the difference between the hops down (from higher to lower state) and the  hops up (from lower to higher state) and accumulated over time. While all three show the same trend up to about 15 fs, SHXF shows almost twice as many D$_2$-D$_0$ hops between about 15 fs and 30 fs than SH and SHEDC, consistent with the observed population dynamics.

\begin{widetext}
	\begin{center}
\begin{table}[h]
	\begin{tabular}{cccccccccccc}
\hline

	 {\bf Mode} &$\omega$ & $\kappa^0$ & $\kappa^1$ & $\kappa^2$ &$\kappa^3$ & $\lambda^{02}$ & $\lambda^{13}$ & $\gamma^0$ & $\gamma^1$ & $\gamma^2$ & $\gamma^3$ \\ \hline
\textbf{18}  & 1193       & -0.02203    & 0.09074     & 0.02748     & -0.04054    & -0.03538     & 0.08077      & 0.01938     & 0.00694     & -0.00294    & 0.00752     \\
\textbf{20}  & 1383       & -0.12147    & 0.05316     & 0.11233     & 0.00747     & -0.02049     &              & 0.01489     & 0.00828     & 0.00183     & 0.00546     \\
\textbf{21}  & 1406       & -0.09468    & 0.04454     & 0.14539     & 0.00050     &              & 0.07284      & 0.00970     & 0.00096     & -0.00114    & 0.01108     
\end{tabular}  
	\caption{Fitted parameters for states D$_0$-D$_3$($\alpha=0-3$) given in eV and frequencies $\omega_{i}$ in cm$^{-1}$ for modes $i=18, 19$ and $20$, taken from Ref.~\cite{AKM15}. Absence of an entry means it is zero.}
	
	\label{table_1}
\end{table}
	\end{center}

\end{widetext}

\begin{widetext}
	\begin{center}
\begin{table}[h]
\begin{tabular}{cccccccccc}
\cline{1-7} \cline{9-10}
	\multicolumn{7}{c}{\textbf{Morse potentials}}                                                                                                                                                                                                             & \multicolumn{1}{c}{} & \multicolumn{2}{c}{\textbf{Coupling constants}}                             \\ \cline{1-7} \cline{9-10}
	\textbf{Mode} & \textbf{State} & $\omega$ & $d_0$ & $a$ & $q_0$ & $e_0$ &                      & $\lambda^{02}$ & $\lambda^{13}$ \\ \cline{1-7} \cline{9-10}
\textbf{24}            &   & 1673 & &  &  & & &    &          -0.18132          
\\
   & D$_0$                                     &                              & 41.89704                            & -0.04719                           & 0.81440                             & 0.81440                             &                      &                                      &                              \\
                                              & D$_1$                                     &                                    & 38.37122                            & -0.05231                           & 0.37488                             & 0.37488                             &                      &                                      &                                      \\
                                              & D$_2$                                     &                                    & 39.25691                            & -0.05286                           & 0.14859                             & 0.14859                             &                      &                                      &                                      \\
                                              & D$_3$                                     &                                    & 37.97847                            & -0.05431                           & -0.18152                            & -0.18152                            &                      &                                      &                                      \\
\textbf{25}     & & 1761          & &  &  & & &  0.00114  &  0.12606                   
\\
& D$_0$                                     &                                & 4.80270                             & 0.13675                            & 0.02883                             & -0.00007                            &                      &                               &                              \\
                                              & D$_1$                                     &                                    & 74.15995                            & 0.03064                            & -1.34468                            & -0.12082                            &                      &                                      &                                      \\
                                              & D$_2$                                     &                                    & 90.76928                            & 0.03374                            & -0.29923                            & -0.00916                            &                      &                                      &                                      \\
                                              & D$_3$                                     &                                    & 20.56979                            & 0.08044                            & 0.38841                             & -0.02071                            &                      &                                      &                                      \\
\textbf{26}        & & 1794  & &  &  & & &  0.13035  &  0.14272  
\\                           & D$_0$                                     &                                & 22.92802                            & -0.07438                           & -0.32069                            & -0.01274                            &                      &                               &                              \\
                                              & D$_1$                                     &                                    & 18.27440                            & -0.07911                           & -0.01711                            & -0.00003                            &                      &                                      &                                      \\
                                              & D$_2$                                     &                                    & 9.46894                             & -0.08653                           & 0.37635                             & -0.01037                            &                      &                                      &                                      \\
                                              & D$_3$                                     &                                    & 65.09678                            & -0.03660                           & 1.66312                             & -0.25639                            &                      &                                      &
\end{tabular}
	\caption{Fitted parameters for states D$_0$-D$_3$ and frequencies $\omega_{i}$ for modes $i=24,25$ and $26$,  taken from Ref.~\cite{AKM15}. The parameters $d_{0}$, $\lambda_{i}^{\alpha\beta}$ and $e_{0}^{\alpha}$ are given in eV and $\omega_i$ in cm$^{-1}$ while $a^{\alpha}$ and $q_{0}^{\alpha}$ are dimensionless.}
	\label{table_2}
\end{table}
	\end{center}
\end{widetext}

\begin{table}[h]
\begin{tabular}{cccc}
\hline
	\textbf{Mode} & $\omega$ & {$\lambda^{01}$} & {$\lambda^{12}$} \\ \hline
\textbf{10}   & 734         & 0.04633      & 0.03148      \\
\textbf{12}   & 771         & 0.03540      & 0.03607     
\end{tabular}
	\caption{Fitted frequencies $\omega_{i}$ and coupling parameters for modes $i=10, 12$,  taken from Ref.~\cite{AKM15}. $\lambda_i^{\alpha\beta}$ are given in eV and $\omega_i$ in cm$^{-1}$.}
	\label{table_3}
\end{table}

\begin{figure}[h]
\includegraphics[width=0.5\textwidth]{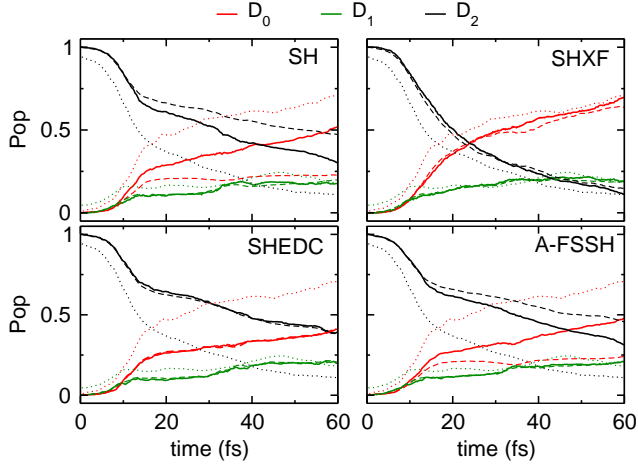}
\caption{Population dynamics for the uracil cation computed from SH, SHEDC, A-FSSH and SHXF along with the reference MCTDH results (from Ref.\cite{AKM15}). The four panels demonstrates the internal consistency of the surface-hopping methods, with the solid lines showing the fraction of trajectories on each state $k$, $\Pi_k(t)$, dashed lines represent the electronic populations, $\rho_{kk}(t)$ and dotted lines are the adiabatic populations from MCTDH calculations}
\label{fig_Uracil_2}
\end{figure}

\section{Supporting Information: Full dimensional vs. LVC model calculations}
\begin{figure}[h]
\includegraphics[width=0.5\textwidth]{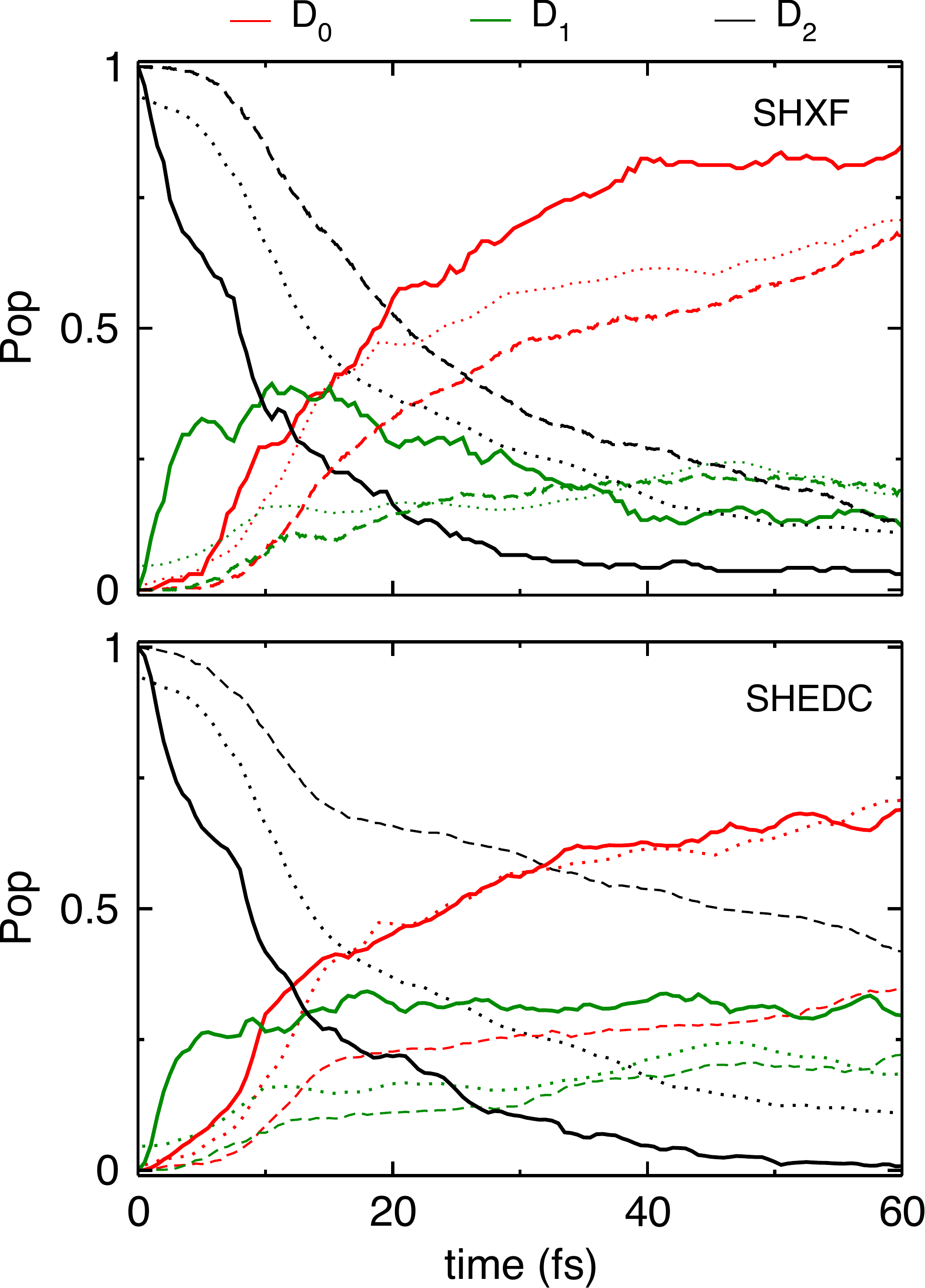}
	\caption{Populations dynamics in the uracil cation computed from SHXF(top) and SHEDC(bottom): full dimensional calculations(solid lines) compared with LVC model results (dashed lines). Dotted lines represent the MCTDH populations,  taken from Ref.~\cite{AKM15}.}
\label{fig_full_LVC}
\end{figure}

Full dimensional SHXF calculations were performed to compare with SHEDC results in Ref.~\cite{AWM16} and those obtained using the LVC model. Since the SHEDC calculations of Ref.~\cite{AWM16} were performed with rescaling along the momentum (isotropic), we also  show the results on the LVC model with this same momentum rescaling (the results in the main text are with scaling along the non-adiabatic coupling vector, but in fact are very similar for this case). 
A nuclear time step of $dt=0.5$fs and the same sigma as in the LVC model were used ($\sigma=0.08$a.u.). In fig.~\ref{fig_full_LVC} we plot the population dynamics from the full dimensional calculations along with the results from the LVC model and MCTDH. The top panel shows the SHXF results and bottom panel SHEDC simulations. Although the overall population trend in the full dimensional case is somewhat similar to MCTDH, the initial behavior is very different. The initial decay of D$_2$ state immediately shows a fast exponential trend, in contrast with the much gentler decay  in the LVC model. Additionally, the initial simultaneous rise of D$_1$ and D$_0$ observed in the eight-mode model is not manifest in the full dimensional calculation. This suggests that, although the three-state intersection certainly appears in the full-dimensional electronic structure~\cite{M08}, the uracil cation does not encounter this region when all the dimensions are considered in the dynamics. (It is interesting that the D$_0$ population from SHEDC on the full-dimensional cation agrees much more closely to that of MCTDH in the reduced dimensional eight-mode model, than it does on the eight-mode model; this is likely just a coincidence as there is no reason it should be).

\begin{figure}[h]
\includegraphics[width=0.5\textwidth]{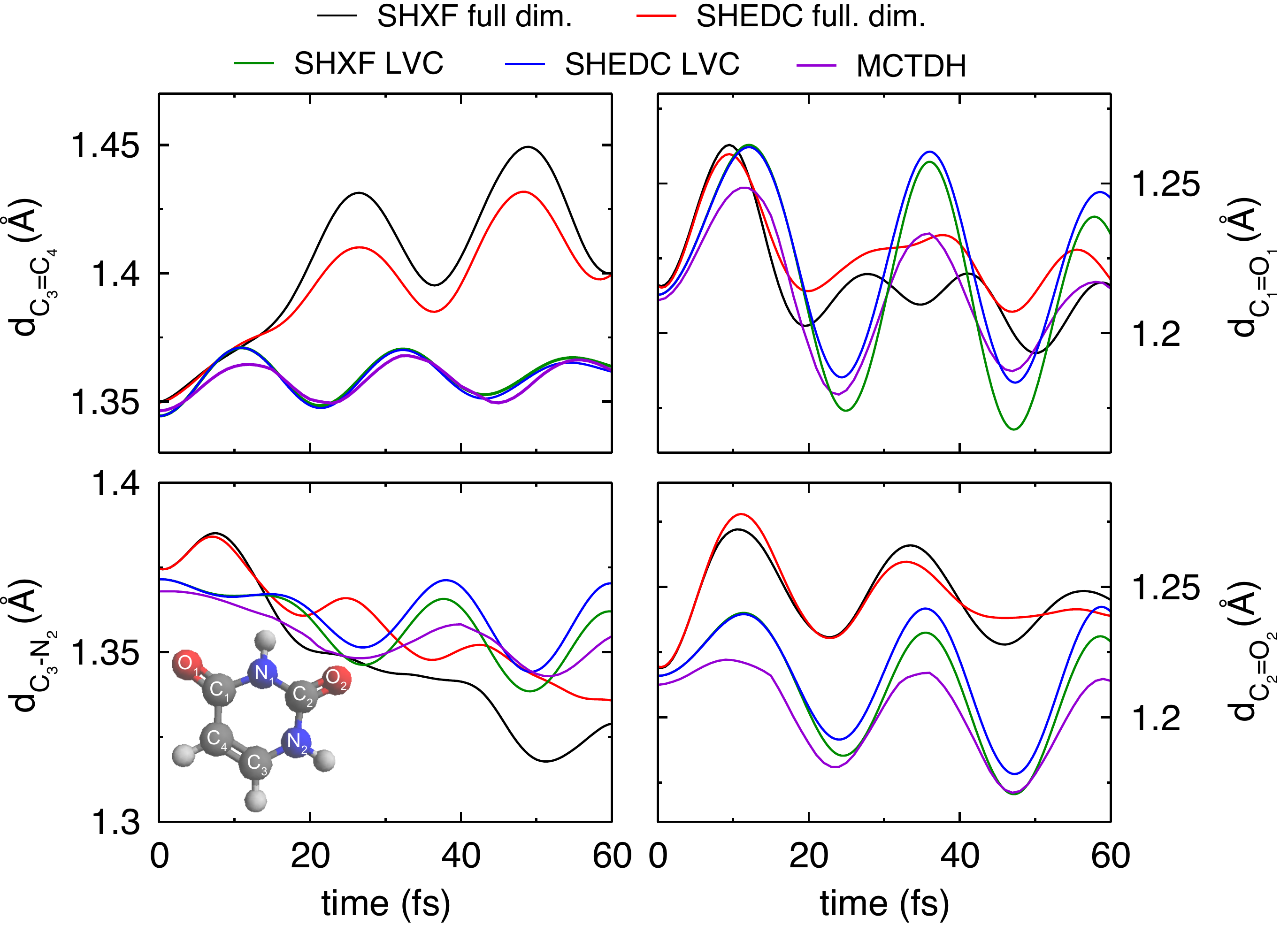}
\caption{Time variation of the uracil cation bond lengths for full dimensional calculations and the LVC model from SHXF (black and green, respectively) and SHEDC calculations (red and blue) and the reference MCTDH (purple): (a) CC double bond, (b) and (c) C=O bonds and (d) C-N bond. }
\label{fig_bonds}
\end{figure}

Figure~\ref{fig_bonds} shows the bond lengths in the different calculations. These bond lengths were chosen as they are related to the important modes included in the LVC model. To correctly capture the bond length dynamics, the harmonic and quartic terms from the $a''$ modes were included in the calculations, in addition to the couplings. The corresponding fitted constants (Eq.~(S7)) are $k^{(0)}_{10}=0.03317$, $k^{(1)}_{10}=0.01157$ and $k^{(1)}_{10}=0.01157$, for mode 10, and  $k^{(0)}_{12}=0.02979$, $k^{(1)}_{12}=0.01488$ and $k^{(2)}_{12}=0.01671$, for mode 12, given in eV (taken from Ref.~\cite{AKM15}). 
Adding these terms in our simulations do not alter the population dynamics, but they are needed for an accurate representation of the motions along the direction of $a''$ modes. When these terms are excluded, we find a larger stretching (not shown)  which ultimately leads to bond breaking.
One can clearly see that, from very short times, the molecule travels through different regions in configurational space in the full-dimensional case as compared to the eight-mode model.  It is worth noting that, within the same model, the trends in the SHXF and SHEDC bond-lengths are similar over this time range. More important, is that our LVC calculations follow quite closely the bond length dynamics predicted by MCTDH, which confirms that SHXF agrees with reference quantum dynamics calculations. 

\clearpage
\bibliography{./ref_na.bib}

\end{document}